\documentclass[aps,twocolumn,groupedaddress,prb,floatfix,showpacs]{revtex4-1}
\usepackage{graphicx}
\usepackage{dcolumn}
\usepackage{bm}
\usepackage{amsmath}
\usepackage{verbatim}

\begin{document}

\title{Density functional theory based study of graphene and dielectric oxide interfaces}

\author{Priyamvada Jadaun}
\email{priyamvada@mail.utexas.edu}
\author{Sanjay K. Banerjee}
\author{Leonard F. Register}
\author{Bhagawan Sahu}
\affiliation{Microelectronics Research Center, The University of Texas at Austin, Austin Texas 78758}
\date{\today}

\begin{abstract}
We study the effects of insulating oxides in their crystalline forms on the energy band structure of monolayer and bilayer graphene using a \textit{first principles} density functional theory based electronic structure method and a local density approximation. We consider the dielectric oxides, SiO$_{2}$ ($\alpha$-quartz) and Al$_{2}$O$_{3}$ (alumina or $\alpha$-sapphire) each with two surface terminations. Our study suggests that atomic relaxations and resulting equilibrium separations play a critical role in perturbing the linear band structure of graphene in contrast to the less critical role played by dangling bonds that result from cleaving the crystal in a particular direction. We also see that with the addition of a second graphene layer, the Dirac cone is restored for the quartz surface terminations. Alumina needs more than two graphene layers to preserve the Dirac cone. Our results are at best semi-quantitative for the common amorphous forms of the oxides considered. However, crystalline oxides for which our results are quantitative provide an interesting option for graphene based electronics, particularly in light of recent experiments on graphene with crystalline dielectrics (hexagonal BN) that find considerable improvements in transport properties as compared to the those with amorphous dielectrics.
\end{abstract}

\maketitle

\section{Introduction}

Recent advances in large area graphene films on metal substrates and understanding the morphology of its domains\cite{rod}, deposition of dielectrics\cite{tutuc}, making metal contacts\cite{vogel} and designing novel electronic switches\cite{sanjay} are making graphene based electronics more likely. Theoretical efforts have played a crucial role in this development often elucidating the complex interplay of substrates and dielectrics with graphene. Notable among these efforts is the understanding of the role of interface and substrate charge inhomogeneity in degrading the carrier mobilities\cite{dassarma}. Density functional theory (DFT) based electronic structure methods, used in some of these studies, have provided inputs for optimization of various processes used in graphene electronics. A crucial obstacle in graphene electronics research has been the need to improve the substrate quality to maintain the novel intrinsic graphene properties. In this article, we address the interface electronic structure of graphene with two crystalline oxides namely, quartz (SiO$_2$) and alumina (Al$_2$O$_3$), using {\it ab-initio} DFT based numerical method. Crystalline substrates/dielectrics provide a compelling alternative to amorphous substrates/dielectrics as they can provide an order of magnitude improvement in the electronic transport properties compared to the the amorphous oxides\cite{philip}. Our studies have implications in interpreting future experiments with crystalline oxides.    

The atomic structure of graphene on amorphous insulating substrates has been explored in some experimental studies\cite{fuhrer} using nanometer scale microscopic techniques such as scanning tunneling microscopy and atomic force microscopy. These studies suggest that graphene conforms to the structural inhomogeneity of the underlying substrate. In addition, a study using Raman spectroscopy examined the role of amorphous oxides on phonon dispersions, indirectly hinting at the change in the electronic spectrum of graphene due to the presence of oxide substrates \cite{balandin}. We are aware of two recent DFT studies of monolayer graphene on a crystalline SiO$_{2}$\cite{theory, nayak}, which discuss perturbations to the linear band structure due to the presence of oxides, the role of dangling bonds and their passivation in influencing the linear band structure.  

\begin{figure}[ht!]
\scalebox{0.18}{\includegraphics{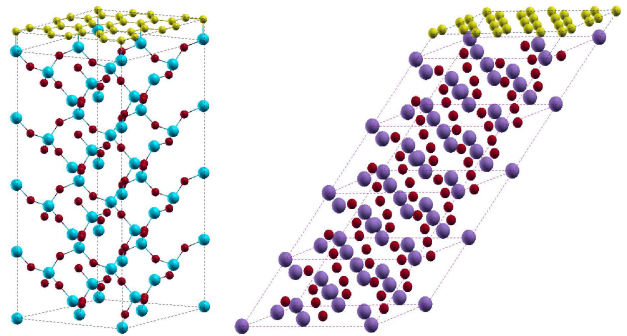}}
\caption{(Color online) Schematic illustrations of supercell structure of monolayer graphene on crystalline oxides. (a) Graphene on Si-terminated quartz with four unit-cells. (b) Graphene on Al-terminated alumina with four unit-cells. The atoms are shown in color: Si(blue), Al(purple), O(red) and C(yellow).}
\label{fig:Fig1}
\end{figure}

Our paper is organized as follows. In section II, we discuss the computational method and convergence parameters used for this study, followed by the motivation and procedure used to build surface models from their bulk counterparts. The results from these calculations are discussed in section III, which details the role of atomic relaxations, the resulting equilibrium distances and the role of unsaturated or dangling bonds of oxide surfaces in perturbing the linear band structure of graphene. We address the origin of change in linear band dispersions with the help of orbital and atom projected densities of states and also present the energetics of our interface models. Finally, we summarize our results and present our conclusions.

\section{Computational Method and Surface Models} 
This section addresses the details of the computational method we used, followed by the procedure adopted to obtain the surface models of quartz and alumina from the bulk and the convergence parameters employed. We adopted a plane-wave based electronic structure method\cite{kresse} using a local density approximation (LDA)\cite{ca} for exchange and correlation, and the projector augmented plane-wave potential for electron-ion interaction\cite{kresse1}. The bulk structures for both quartz and alumina are consistent with those mentioned in literature\cite{poindexter} and the surface models built from the bulk structures conform to the widely accepted quartz and alumina structures\cite{somorjai} containing alternating cation and anion layers.

The unit cell of bulk quartz contains 27 atoms with 3 silicon planes and 6 oxygen planes, each plane comprising three atoms. The unit cell of alumina contains 30 atoms with 4 aluminium planes and 6 oxygen planes, again with each plane consisting of only three atoms. We used a 7 $\times$ 7 $\times$ 5 \textbf{k}-point mesh in the hexagonal Brillouin zone (BZ) and a kinetic energy cut-off of 612 eV. The results were carefully checked in comparision to a larger \textbf{k}-point set and higher energy cut-offs.

\begin{figure}
\scalebox{0.3}{\includegraphics{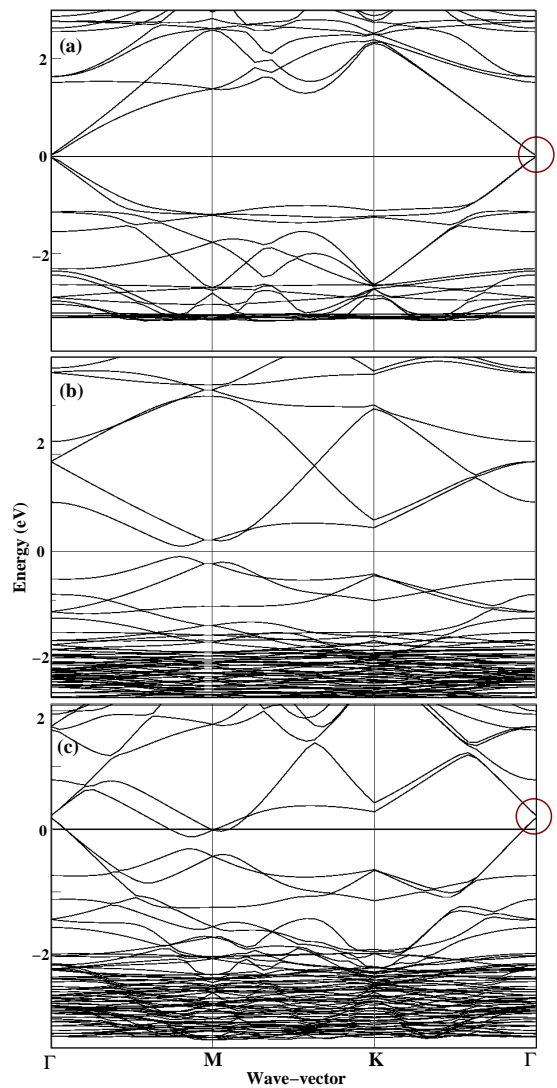}}
\caption{Energy band structures of (a) monolayer graphene on Si-terminated quartz (b) monolayer graphene on O-terminated quartz (c) bilayer graphene on O-terminated quartz. The atoms in the supercell were relaxed. The Fermi energy is set at zero. The origin of occurrence of linear bands at the $\bf \it \Gamma$-point instead of {\bf \it K}-point of the supercell is explained in the text.}
\label{fig:Fig2}
\end{figure}

After optimization of atomic positions and lattice constant for the bulk structures, we found these values to be very close to the experimental values, differing by less than 1$\%$ (Table I). Using these optimized lattice parameters, we constructed surface models of both quartz and alumina. Four oxide bulk unit cells were chosen to represent the thin-films and were stacked along the crystallographic {\it c}-direction (corresponding to a film thickness of about 22 \text{\AA}). At their top termination, we placed a graphene layer. We found that 6 $\times$ d$_{C-C}$ graphene, containing 24 carbon atoms (where d$_{C-C}$ = 1.42 \text{\AA}), was nearly commensurate with the hexagonal surface of the oxides (Fig. 1). With this set-up, we get the lattice mismatch of quartz and alumina with graphene as $\sim$ 0.19$\%$ and $\sim$ 0.42$\%$, respectively. The lattice mismatch values mentioned here serve only as an initial guideline towards assessing the degree of distortion of graphene in presence of the oxides. We keep these mismatched values of lattice constants by not allowing any variation in supercell size.
These mismatched values result in a weak rippling effect of carbon atoms which may not be observed in experiments using graphene on atomically flat substrates. We estimate average height fluctuations of relaxed two-dimensional graphene and find that the deviations are approximately 0.05 \text{\AA}, which is small compared to fluctuations in suspended graphene \cite{kats}. 

Modeling roughness present on amorphous oxide surfaces by DFT is challenging partly due to the inherent complexity in generating the amorphous structures by DFT (the traditional melt and quench method) as well as the requirement of relatively large size of the supercell to represent roughness, a computationally intensive task for DFT calculations. Moreover, the amorphous oxide structures do not possess dangling bonds. Therfore, an additional constraint in using the crystalline substrates, besides treating surface roughness in an approximate way, is to saturate these dangling states for assessing the external effects on graphene electronic structure. To make the combined graphene and oxide thin films structure an isolated system, we used a vacuum region adjacent to graphene layer in the stacking direction and supercell is then repeated in all three crystallographic direction by imposing periodic boundary conditions on the crystal wave functions. 
The dangling states at the bottom of the supercell were saturated with hydrogen atoms and the atoms within the top two unit-cells of the oxide film and all the graphene atoms were allowed to relax while keeping the cell size fixed. We used the same energy cut-off value as in the bulk calculations, but the \textbf{k}-point mesh in the BZ was chosen to be 7 $\times$ 7 $\times$ 1. The convergence of results with respect to larger {\bf k}-mesh size, larger vacuum sizes and energy cut-off was also checked. For atomic relaxation, the total energy was assumed to have converged when all the components of the Hellman-Feynman forces were smaller than 0.01 eV/\text{\AA}.

\begin{figure}[ht]
\scalebox{0.3}{\includegraphics{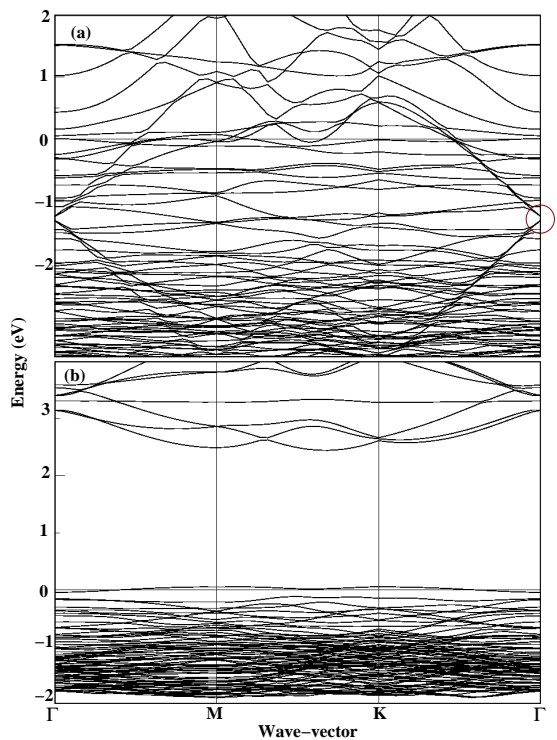}}
\caption{Energy band structures of (a) monolayer graphene on Al-terminated alumina (b) monolayer graphene on O-terminated alumina. The atoms in the supercell were relaxed. The Fermi energy is set at zero.}
\label{fig:Fig3}
\end{figure}

\section{Energy Band dispersions, Densities of states and Energetics}

In this section, we discuss the effect of underlying oxide surfaces on the electronic band structure of graphene for different surface terminations. In particular we explore the impact of graphene-oxide distances, oxide dangling bonds and atomic relaxations on the band structure.  With relaxations of atomic positions, we get equilibrium oxide-graphene distances tabulated in Table I. All the distances are  larger than the corresponding interatomic distances of the constitutent atoms in bulk or molecular compound phases, suggesting that formation of compound phases at low temperature is unlikely. However, higher temperture effects may support formation of these phases. 
We present detailed examination for all four surface terminations, beginning with silicon-terminated quartz. 
Figure 2(a) shows the self-consistent band structure of graphene on silicon-terminated quartz at the equilibrium distance (Table I). It
is evident that linear band structure is not perturbed at such a distance, which is consistent with atom and orbital projected density of states (DOS) (figure not shown). We find weak hybridization between carbon-{\it p} and silicon-{\it p} orbitals in the valence band region up to the Fermi level. We note here that the Dirac point occurs at the center of the supercell BZ and not at the {\it K} point of the honeycomb BZ due to symmetry reasons explained in the Appendix\cite{farjam}. For silicon-terminated quartz, the graphene sheet remains flat. On oxygen-terminated quartz, our calculations suggest noticeable rippling of graphene. Moreover, the small equilibrium interface distance of 1.76 \text{\AA} (Table I) leads to perturbation in graphene's band structure (Fig. 2(b)). The DOS plot (Fig. 4(a)) shows considerable hybridization in the whole valence band region between oxygen-{\it p} and carbon-{\it p} orbitals. A similar conclusion was reached in a recent DFT-based calculation of monolayer graphene on oxygen-terminated quartz\cite{nayak}. Interestingly, adding one more layer of graphene to the oxygen-graphene supercell allows the first layer to act as a buffer, protecting the Dirac cone (Fig. 2(c)). Both surface terminations of alumina result in strong perturbations of the graphene Dirac cone (Fig. 3(a) and (b)) which is consistent with the DOS plots (Fig. 4(b) and 4(c)). We note that adding an extra layer of graphene to the alumina surface terminations does not recover the linear band structure, suggesting more than two graphene layers are needed to protect the Dirac cone.     

To understand the distance-dependent electronic structure of the combined graphene and oxide system, we manually adjusted the interface distances both below and above the calculated equilibrium distances. The atomic positions were not relaxed while other computational parameters kept the same as those used in the self-consistent runs discussed above. We demonstrate our calculations by using silicon-terminated quartz as an example. We chose two representative distances 2 \text{\AA} and 4 \text{\AA} which are, respectively, below and above the equilibrium distance of 3 \text{\AA} (Table I). The Dirac cone is protected in 4 \text{\AA} case (Fig. 5(a)) but not in 2 \text{\AA} case (Fig. 5(b)), suggesting a strong dependence of band structure on the interfacial distance. We have performed these tests with other surface terminations and found that the interfacial distance dictates whether the Dirac cone can be protected or not. The equilibrium distance after relaxation is thus quite important in determining graphene's properties.

\begin{figure}[ht]
\scalebox{0.4}{\includegraphics{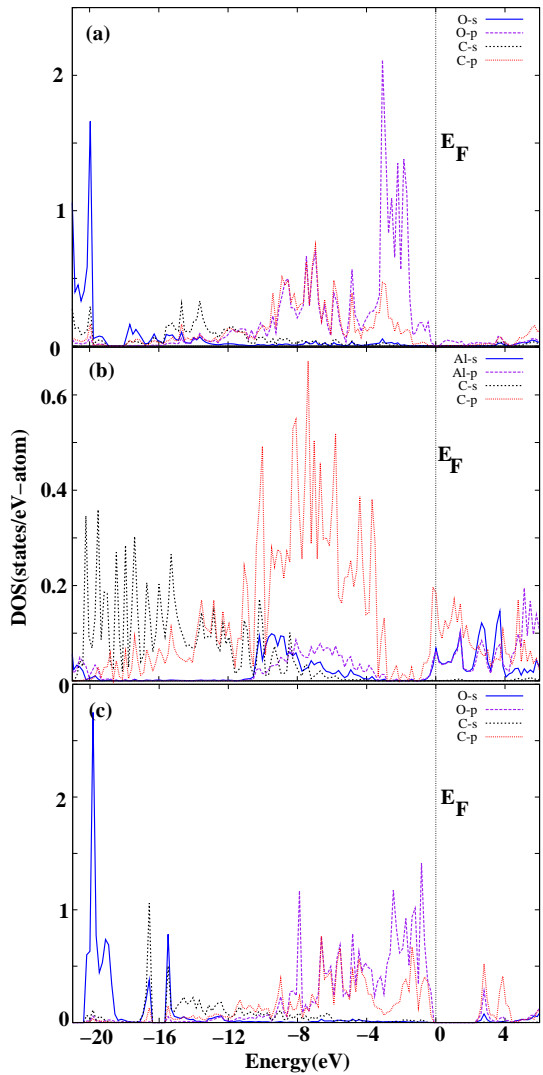}}
\caption{ (Color online) Atom and orbital projected density of states for a C atom in monolayer graphene and the atoms of the topmost planes in (a) O-terminated quartz (b) Al-terminated alumina  and (c) O-terminated alumina oxides. The atoms in the supercell were relaxed. The Fermi energy is set at zero.}  
\label{fig:Fig4}
\end{figure}

Table I shows the binding energy values for graphene on four surface terminations considered in this study. In the most relevant case, i.e., no surface passivation but with atomic relaxation, these values hint at the non-bonding nature of graphene to the underlying oxide oxides. The binding energy values are obtained by using the following definition.

\begin{equation}
 {\rm {\it E}^{bind} = {\it E}(supercell)- {\it E}(Gr) - {\it E}(oxide)}
\end{equation}

where {\it E}(supercell) denotes the total energy of the supercell containing the oxide and a graphene layer. {\it E}(Gr) and {\it E}(oxide) denote, respectively, the total energies of isolated graphene and isolated oxide in the same supercell set-up, with the same energy cut-off and {\bf k}-point mesh as that of the combined graphene and oxide calculations.

\begin{table}[ht]
\caption{
Lattice parameters (in \text{\AA}) of bulk quartz and alumina, average interplanar distances (in \text{\AA}) of graphene from the four underlying surface terminations and their binding energies (in eV/atom), in case of non-passivated but relaxed surfaces. The numbers in parenthesis are the out-of-plane lattice constants of the bulk phases, and those in square parenthesis are interatomic distances between carbon and the corresponding atom of the surface terminations in the bulk or molecular phase. It should be noted that for isolated oxide calculations we passivated the top layer with hydrogen atoms to keep the supercell non-magnetic.  
}
\label{tab1}
\begin{ruledtabular}
\begin{tabular}{cccccccc}
                       &  Lattice Paramters & d(C-x)                & E$^{bind}$       \\ 
                       &                     & (x = Si,Al,O)         &                    \\ \hline
Bulk quartz            &                    &                       &                   \\
 This work             &  4.914 (5.408)     &                       &                   \\
 Expt.\footnote{Reference 17, 21}    &  4.913 (5.405)     &                   &                   \\
Si-terminated           &                    &    3.0 (1.89)\footnote{Reference 22}         & 10.061            \\
O-terminated           &                    &    1.76(1.3)\footnote{Reference 23}         & 0.581            \\
                       &                    &                       &                     \\ \hline
Bulk alumina          &                    &                       &                     \\
This work              & 4.907 (4.908)      &                       &                     \\
Expt.\footnote{Reference 18, 24}     & 4.943 (4.907) &                       &                     \\
Al-terminated          &                    &   2.7 (1.89-2.19)\footnote{Reference 25}     &   1.017      \\
O-terminated          &                    &   2.15                &   0.689              \\
\end{tabular}
\end{ruledtabular}
\end{table}

\begin{figure}[ht]
\scalebox{0.3}{\includegraphics{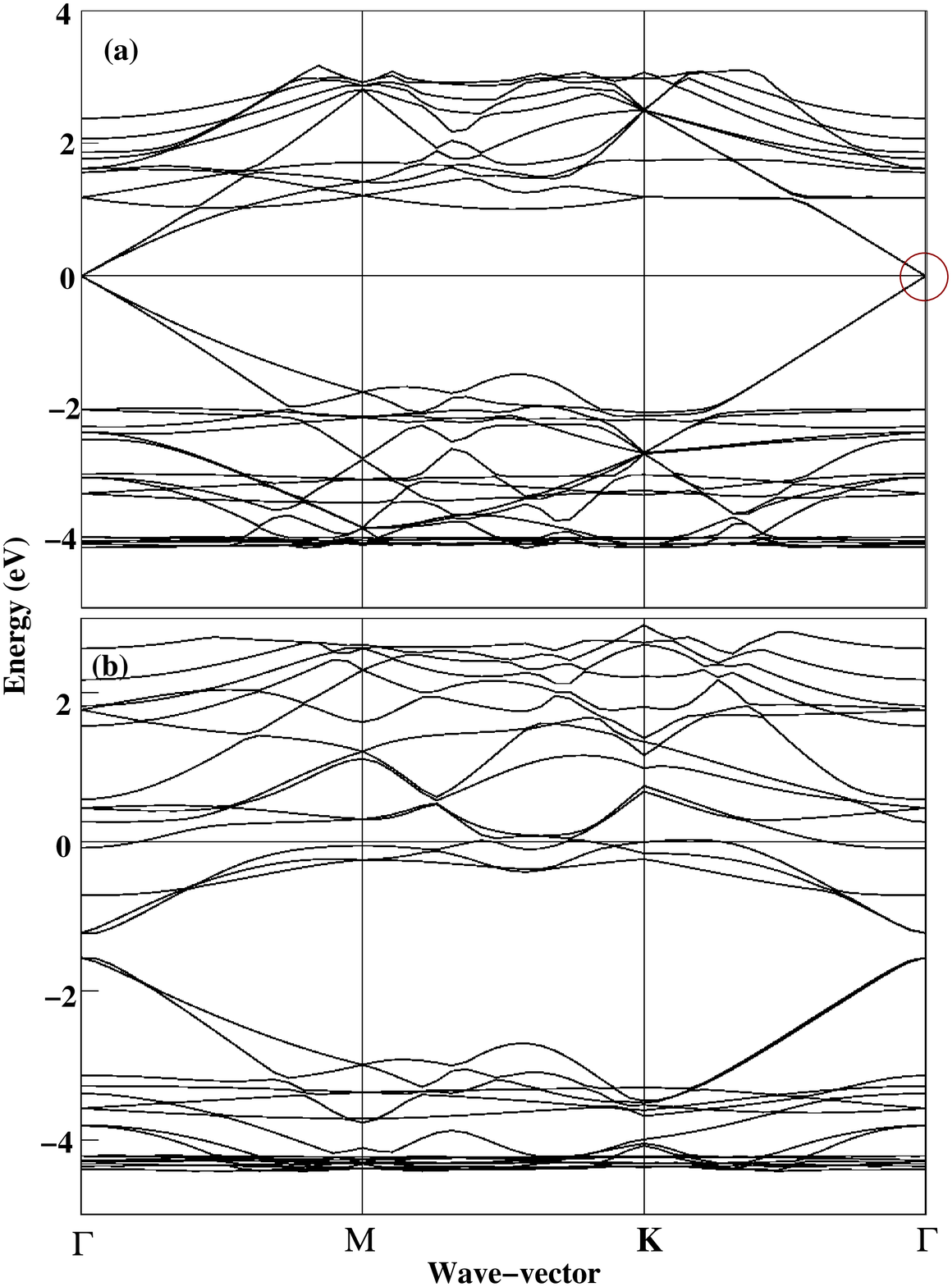}}
\caption{Energy band structures of (a) monolayer graphene on Si-terminated quartz at a distace of 4 \text{\AA} (b) monolayer graphene on Si-terminated quartz at a distance of 2 \text{\AA}. The atoms in the supercell were not relaxed. The Fermi energy is set at zero. The oxide surface was unpassivated.}
\label{fig:Fig5}
\end{figure}

To assess the importance of dangling bond saturations on graphene's electronic specturm, we performed a calculation with the top oxide surface saturated with hydrogen atoms. This calculation was undertaken only for oxygen-terminated surface of quartz since silicon-terminated surface already protects the Dirac cone without surface passivation. Both aluminium- and oxygen-terminated surfaces of alumina, even in the initial structures, show non-trivial stoichiometry, as a result passivation of dangling bonds is avoided. Two separate calculations were performed with equilibrium oxygen-carbon interfacial distance in oxygen-terminated quartz: one with un-relaxed and unpassivated oxide surfaces, and another with un-relaxed passivated oxide surface with graphene on top of it. We find that on un-passivated surfaces, there is no effect on graphene's band structure except for some doping effects (Fig. 6). This hints at the non-critical role played by dangling bond states. 
To check the effect of relaxation of the unpassivated surfaces in all four terminations without the graphene layer, we performed additional calculations which show that both aluminum- and silicon-terminated surfaces are inherently stable compared to the initial surface model we built from the corresponding bulk structures as well as the final relaxed unpassivated structures. The interatomic and interplanar distances were found to be not different from the unrelaxed initial and relaxed final structures and the surface atomic configurations were similar. This situation corresponds to the configuration adopted by the silicon and aluminium surface terminations in experiments immediately after cleaving from the bulk crystals. Therefore, our initial surface models with graphene on the top of these surfaces and the corresponding interfacial distances in Table I obtained upon relaxing these combined structures without performing the intermediate step relaxing the unpassivated oxide surfaces, are valid. The situation is not same with oxygen-terminated surfaces of quartz and alumina. Both surfaces show large relaxations of in-plane surface atoms. However, they are all displaced by the same amount. To a first order approximation, this uniform displacement is equivalent to displacing the graphene layer in-plane with respect to the oxide surface. Since our graphene position is an independent variable, this displacement would not matter. We believe all second-order effects of relative in-plane movement between substrate layers are taken care of in the relaxations of the combined structures that we perform.            

\begin{figure}[ht]
\scalebox{0.3}{\includegraphics{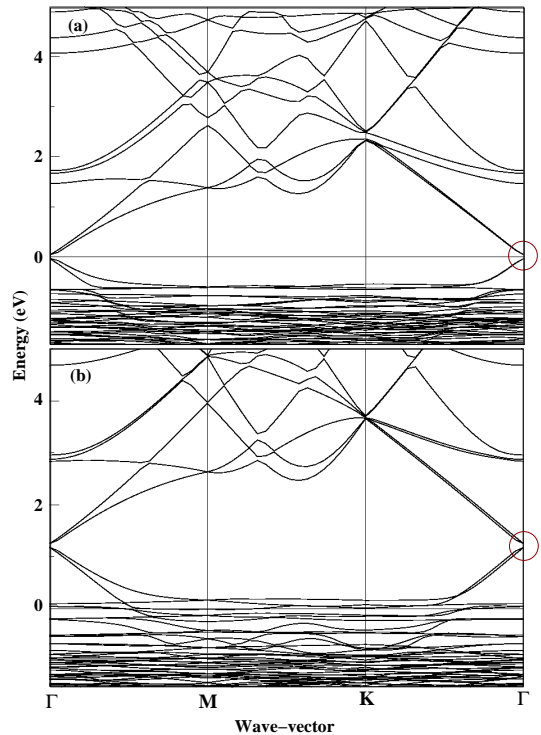}}
\caption{Energy band structures of (a) monolayer graphene on O-terminated quartz at a distace of 1.76 \text{\AA}. The top surface of oxide is passivated with hydrogen atoms (b) monolayer graphene on O-terminated quartz at a distance of 1.76 \text{\AA}. The top surface of oxide is not passivated in this case. The atoms in the supercell were not relaxed. The Fermi energy is set at zero.}
\label{fig:Fig6}
\end{figure}

\section{Summary and Conclusion} 

In summary, we have studied, using a {\it first principles} DFT method, the effect of two crystalline insulating oxides, quartz and alumina, on the electronic structure of monolayer graphene. Silicon-terminated quartz retains the graphene band structure on non-passivated surface with atomic relaxations. Oxygen-terminated quartz and both alumina terminations perturb the linear bands of graphene. Graphene and oxide interface distance is critical in determining the extent of disturbance to graphene. By comparison, the effect of dangling bond states on the oxide surface is not very crucial. Bilayer graphene on oxygen-terminated quartz retrieves its linear band structure while graphene on alumina terminations (Al or O) perhaps needs additional graphene layers. Our studies are relevant in situations where a crystalline dielectric is used either as a template for growth or as a substrate/dielectric for exfoliated graphene experiments. Recent use of crystalline BN as a substrate/dielectric for graphene experiments is a promising step in this direction, and we believe more experiments with crystalline oxides should be conducted in future, which make DFT-based studies useful for interpreting experiments.

\acknowledgments

The authors acknowledge financial support from the DARPA-CERA and NRI-SWAN program. The authors acknowledge the allocation of computing time on NSF Teragrid machines {\it Ranger} (TG-DMR080016N) and {\it Lonestar} at Texas Advanced Computing Center.    
\appendix*
\section{}

Quartz, alumina and the graphene layer, all have a hexagonal unit cell. Let $\roarrow A_{1}$, $\roarrow A_{2}$ and $\roarrow A_{3}$ be the lattice vectors of the entire graphene and oxide system and $\roarrow B_{1}$, $\roarrow B_{2}$, and $\roarrow B_{3}$ be the correponding reciprocal lattice vectors. Similarily, let ($\roarrow a_{1}$, $\roarrow a_{2}$ , $\roarrow a_{3}$) and ($\roarrow b_{1}$, $\roarrow b_{2}$ and $\roarrow b_{3}$) be the triad of primitive and reciprocal lattice vectors of the graphene layer alone.

We note that the lattice structure of the graphene plus oxide system is a (2 $\sqrt(3)$ x 2 $\sqrt(3)$)R30$^o$ reconstruction of the lattice structure of the graphene layer alone.
Thus $A = 2 \sqrt(3) a$ where A is the lattice constant of the entire system and a is the lattice constant of the two-dimensional graphene layer. Since both unit cells are hexagonal this implies that the reciprocal lattice structure of the graphene layer on top is a ($2\sqrt(3)$ x $2\sqrt(3)$)R30$^o$ reconstruction of the reciprocal lattice structure of the graphene plus oxide system below. Taking $\roarrow a_{1}$ to be along $\hat x$,

\begin{equation}
 {\rm \roarrow a_{1} = {\it a} \hat x} 
\end{equation} 
\begin{equation}
 {\rm \roarrow a_{2} = {\it a} (\frac{1}{2} \hat x + \frac{\sqrt(3)}{2} \hat y)} 
\end{equation} 
\begin{equation}
 {\rm \roarrow A_{1} = 2 \sqrt(3) {\it a}(\frac{\sqrt(3)}{2} \hat x + \frac{1}{2} \hat y)} 
\end{equation} 
\begin{equation}
 {\rm \roarrow A_{2} = 2 \sqrt(3) {\it a} \hat y} 
\end{equation} 
\begin{equation}
 {\rm \roarrow A_{3} =  {\it c} \hat z} 
\end{equation}

\begin{figure}[ht]
\scalebox{0.35}{\includegraphics{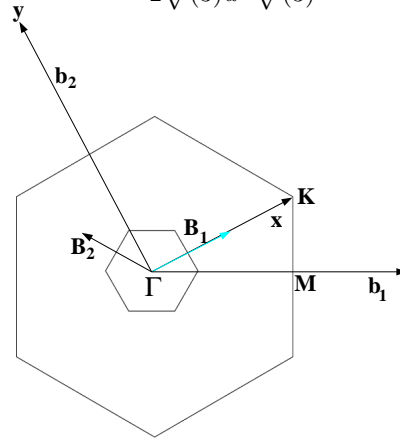}}
\caption{ (Color online) Schematics of supercell Brillouin zone (small) superposed on the graphene honeycomb Brillouin zone (large). The reciprocal lattice vectors are shown along with the overlap of $\bf \Gamma$ to {\bf K} vector of graphene with the {\bf B$_1$} vector of the supercell. The overlap is due to the symmetry reasons and is explained in the text.}   
\label{fig:Fig7}
\end{figure}

The reciprocal lattice vectors are,
 
\begin{equation}
 {\rm \roarrow b_{1} = \frac{2 \pi}{\it a} (\hat x - \frac{1}{\sqrt(3)} \hat y)} 
\end{equation} 
\begin{equation}
 {\rm \roarrow b_{2} = \frac{2 \pi}{\it a} ( \frac{2}{\sqrt(3)} \hat y)} 
\end{equation} 
\begin{equation}
 {\rm \roarrow B_{1} = \frac{2 \pi}{ 2 \sqrt(3) \it a} ( \frac{2}{\sqrt(3)} \hat x)} 
\end{equation} 
\begin{equation}
 {\rm \roarrow B_{2} = \frac{2 \pi}{ 2 \sqrt(3) \it a} ( -\frac{1}{\sqrt(3)} \hat x + \hat y)}
\end{equation} 

Figure 7 shows the BZ set-up with these vectors using equations A.6 to A.9.

Now if we take the same $\Gamma$ point as the center of both the Brillouin zones and draw the reciprocal unit cell of both the structures, we realize that the {\bf K } point of only the graphene layer lies on the reciprocal lattice vector $\roarrow B_{1}$ of the graphene plus oxide system (as shown in figure). This means that the {\bf K } point of graphene layer folds in by symmetry onto the $\Gamma$ point of the entire system.

\end{document}